\documentclass[twocolumn,preprintnumbers,nofootinbib,prd]{revtex4}
\usepackage{graphicx}

\usepackage[hypertex]{hyperref}
\newcommand{\vev}[1]{\langle {#1} \rangle}
\newcommand{\lsim}{\lesssim}
\newcommand{\ord}[1]{\mathcal{O}{(#1)}}

\newcommand{\gsim}{\gtrsim}
\newcommand{\beq}{\begin{equation}}
\newcommand{\eeq}{\end{equation}}

\begin{document}

% page numbers bottom-center
\pagestyle{plain}

\preprint{MADPH-05-1424}

\title{Probing the Origins of Neutrino Mass with Supernova Data}

\author{Hooman Davoudiasl\footnote{email: hooman@physics.wisc.edu}}

\author{Patrick Huber\footnote{email: phuber@physics.wisc.edu}}

\affiliation{Department of Physics, University of Wisconsin,
Madison, WI 53706, USA}

%%%%%%%%%%%%%%%%%%%%%%%%%%%%%%%%%%%%%%%%%%%%%%%%%%%%%%%%%%%%%%%%%%%%%%%%%%%%

\begin{abstract}

We study type II supernova signatures of neutrino mass generation
via symmetry breaking at a scale in the range from keV to MeV. The scalar
responsible for symmetry breaking can be thermalized in the
supernova core and restore the symmetry. The neutrinos from scalar
decays have about half the average energy of thermal neutrinos and
are Bose-Einstein distributed.  We find that, even without a
detailed knowledge of the supernova parameters, a discovery is
well within reach at Super-Kamiokande.

\end{abstract}
\maketitle

%%%%%%%%%%%%%%%%%%%%%%%%%%%%%%%%%%%%%%%%%%%%%%%%%%%%%%%%%%%%%%%%%%%%%%%%%%%

Type II supernov{\ae} provide some of the most extreme
environments at the present cosmological epoch.  In these
explosions, the stellar gravitational binding energy $E_\mathrm{SN} \simeq
3 \times 10^{53}\,\mathrm{erg}$ is released almost entirely in a neutrino
burst. After the initial collapse, the stellar core is
characterized by nuclear densities and a temperature $T_\mathrm{SN} \simeq
30\,\mathrm{MeV}$.  The cooling of the core occurs over a period
of seconds, through an explosive release of $E_\mathrm{SN}$ by neutrino emission.

Given such conditions, it is interesting to ask whether there may
be new particles at a scale below $T_\mathrm{SN}$ that could be
relevant in the first few seconds of the supernova explosion. In
fact, supernova cooling has provided stringent bounds on many
models containing light particles, like the
axion~\cite{Eidelman:2004wy}. However, such models do not
generically yield distinct supernova signatures.

In this Letter, we study possible supernova signatures of neutrino
mass generation from low energy symmetry breaking. The basic idea
is that, given a typical supernova environment with temperature
$T_\mathrm{SN}$, a light scalar, whose vacuum expectation value ({\sc vev})
generates $m_\nu \neq 0$, may be in thermal equilibrium during the
explosion. We show that this can typically be the case,
providing robust supernova signatures of this physics,
which may otherwise be hard to access.

Our study is based on recent low-cutoff-scale models for neutrino
mass generation proposed in Ref.~\cite{Davoudiasl:2005ks}. In
contrast to the seesaw mechanism with scales of order
$10^{14}\,\mathrm{GeV}$, these models can be cut off at $10\,\mathrm{TeV} \lsim
\Lambda \lsim 1000\,\mathrm{TeV}$, avoiding extrapolations far above the
experimental frontier at around $1\,\mathrm{TeV}$. Both Majorana and Dirac
masses can be generated in these theories. To afford a low cutoff
scale, a discrete gauged (anomaly-free) $Z^{\ell}_3 \times Z^q_9$
symmetry group is assumed to protect Baryon ($B$) and Lepton ($L$)
numbers.  Here, small values of $m_\nu \neq 0$ are generated when
certain symmetries are spontaneously broken at or below $10\,\mathrm{MeV}$,
by a scalar {\sc vev}\footnote{Refs.\cite{Chacko:2003dt,Chacko:2004cz}
contain similar ideas, using global symmetries.}.  In this work,
we generically denote such a scalar by $\varphi$.

A consequence of this scenario is that a Bose-Einstein gas of
scalars with $T \simeq T_\mathrm{SN}$ will be present during the cooling of
the supernova. Thermal scalars produced in the bulk are either confined within
the neutrino-sphere or they decay into neutrinos, which are also confined.
The $\varphi$'s escaping at the surface quickly decay into neutrinos.
The average energy of these neutrinos is half that of the thermal scalars.
In addition, these neutrinos inherit the
Bose-Einstein distribution of the parent scalars.
These are distinct signatures of the presence of a light bosonic
degree of freedom that couples to neutrinos. Detecting a flux of
supernova neutrinos with about half the expected average energy and a
Bose-Einstein thermal origin will strongly favor the type of models
we consider.  In this case, the supernova data will provide {\it a
direct probe of the mechanism responsible for neutrino mass
generation.}

Since the thermalized scalars are expected to have a self coupling
$\lambda \simeq 1$, the corrections to their potentials are of order
$\lambda T^2 \gg \vev{\varphi}^2$, where $T \simeq T_\mathrm{SN}$. Thus,
another consequence of these models is that the symmetry whose
breaking gives $m_\nu \neq 0$ is restored in the supernova core.
We then expect that $\vev{\varphi} = m_\nu = 0$ in the core,
during the initial stages of the supernova explosion. Thus a
symmetry breaking {\it phase transition} will take place after the
star has cooled to small enough temperatures, $T \lesssim
\vev{\varphi}$. This will result in the emission of non-thermal
neutrinos with $E_\nu \simeq \vev{\varphi}$. Later, we will briefly
comment on the possibility of detecting such a signal.

To demonstrate these features, we adopt a neutrino mass model with
low scale symmetry breaking, presented in
Ref.~\cite{Davoudiasl:2005ks}. In this model, Majorana neutrino
masses consistent with flavor oscillation data are generated.
Other models with more degrees of freedom that lead to Dirac
masses from low scale symmetry breaking have also been considered
\cite{Davoudiasl:2005ks}. To keep the discussion simple and
emphasize the main features, we will not consider these latter
models here. However, a similar analysis can be performed for the
Dirac mass models of Ref.~\cite{Davoudiasl:2005ks}.

Without specific assumptions about the cutoff scale $\Lambda$, a
gauged $Z^{\ell}_3 \times Z^q_9$ protects $B$ and $L$ numbers at
safe levels \cite{Davoudiasl:2005ks}.  To obtain Majorana masses
for neutrinos, a scalar $\varphi$ with $Z^{\ell}_3$ charge $+1$ is
introduced. Suppressing the generation indices, the dimension-6
operator
\beq
O_\varphi = \frac{\varphi L H L H}{\Lambda^2} \;
\label{d6}
\eeq
then yields the Yukawa coupling $y \,\varphi \, \nu_L
\nu_L$, with $y \equiv (v/\Lambda)^2$ and $v \equiv \vev{H} \neq
0$; $L$ and $H$ are the Standard Model lepton doublet and the Higgs, 
respectively.

The coupling $y$ is constrained by $0 \nu \nu \beta + \varphi$
\cite{Bernatowicz:1992ma}: $y < 3 \times 10^{-5}$, implying that
$\Lambda \gsim 30\,\mathrm{TeV}$. For $m_\nu \simeq 0.1\,\mathrm{eV}$,
we then need
$\vev{\varphi} \gsim 3\,\mathrm{keV}$. On the other hand, bounds on
cosmological domain walls from broken discrete
symmetries
require $\vev{\varphi} \lsim 1\,\mathrm{MeV}$, which yields $\Lambda
\lsim 10^3\,\mathrm{TeV}$.  In this work, we will then consider
\beq
1\,\mathrm{keV} \leq \vev{\varphi} \leq 1\,\mathrm{MeV}\,.
\label{chirange}
\eeq
The upper-bound on
$\vev{\varphi}$ can easily be increased by $\ord{1}$ factors if
the domain wall network is frustrated or if the reheat temperature
in the Early Universe is below the symmetry breaking scale.

In a typical supernova environment, the scalar $\varphi$ will be
thermally produced in equilibrium and, consequently, the broken
symmetry responsible for $m_\nu \neq 0$ will be restored. To see
this, we first note that after the initial collapse of the star,
the hot and dense gas of particles around the core of the
progenitor traps neutrinos, forming a {\it neutrino-sphere}.  The
$\nu$-confinement typically lasts over the cooling time, $\Delta
t_{\rm cool} \sim 10\,\mathrm{s}$, of the supernova.  Assuming that the
neutrinos are in thermal equilibrium inside the neutrino-sphere, their
number density is given by $n_\nu \sim T_\mathrm{SN}^3$. The rate of the
reaction $\nu_L \nu_L \to \varphi$ is given by $\Gamma_\varphi
\simeq y^2 T_\mathrm{SN}$ which, up to phase space factors, is of the same
order as the rate for $\nu_L \nu_L \to \varphi \varphi \varphi$,
given a $\varphi$ 4-point-coupling $\lambda \simeq
1$.\footnote{There is also the process $\nu_L \nu_L \to \varphi
\varphi$ mediated by the coupling $m \varphi^3$.  With a natural
potential for $\varphi$, we have $m \simeq m_\varphi \simeq
\vev{\varphi}$. For $T \gg \vev{\varphi}$, this process is then
sub-dominant to the first two.} From (\ref{chirange}), we expect
$y \gsim 10^{-7}$, with $m_\nu \simeq 0.1\,\mathrm{eV}$.  This implies
$\Gamma_\varphi \gsim (10^{-8}\,\mathrm{s})^{-1}$.

Thus, $\Gamma_\varphi^{-1} \ll \Delta t_{\rm cool}$.  Note that
$\Gamma_\varphi$ also sets the rate for $\nu_L \varphi \to \nu_L$
and $\nu_L \varphi \to \nu_L \varphi \varphi$, mediated by the
$t$-channel exchange of $\varphi$.  The mean-free-path $d_\varphi$
for the interactions of the $\varphi$'s produced inside the
neutrino-sphere is then given by $d_\varphi \sim
\Gamma_\varphi^{-1} \lsim 1\,\mathrm{m}$.  The size $R_\nu$ of the
neutrino-sphere is comparable to that of a neutron-star and we
have $R_\nu \simeq 50 \,\mathrm{km} \gg d_\varphi$. Given all these
considerations, we conclude that the neutrinos and the $\varphi$'s
come to thermal equilibrium inside the neutrino-sphere, well
before the star begins cooling.  This means that the number
density of $\varphi$'s is given by $n_\varphi \sim n_\nu \sim
T^3_\mathrm{SN}$.

To study finite temperature effects, we write down a
$Z^\ell_3$-invariant potential for $\varphi$ below the weak scale
\beq
V(\varphi) = y \varphi \nu_L \nu_L - \mu^2 \varphi
\varphi^\dagger + m \varphi^3 + \lambda (\varphi
\varphi^\dagger)^2 + \mathrm{h.c.}
\label{Vchi}
\eeq

In a typical theory, $\mu \simeq m \simeq \vev{\varphi}$, and $\lambda
\simeq 1$. Since experimental constraints require $y < 3 \times
10^{-5}$, the Yukawa coupling is much smaller than
other typical couplings of the $\varphi$ system. Hence, Yukawa
contributions to thermal corrections of $V(\varphi)$ are ignored
in our analysis.

Once in equilibrium, the $\varphi$ field obtains a thermal mass
$\sim \lambda \, T^2$.  At $T \simeq T_\mathrm{SN}$, this correction is
much larger than the typical negative mass squared responsible for
symmetry breaking: $\lambda \, T_\mathrm{SN}^2 \gg \mu^2$.  Hence, the
spontaneously broken symmetry is restored within the
neutrino-sphere.

The temperature of the neutrino-sphere will eventually fall below
the critical temperature $T_c$ at which the symmetry is broken.
Here, we note that due to the presence of the term $m \varphi^3$,
the transition back to the broken phase may be first order.  Quite
generically, we expect $T_c \simeq \vev{\varphi}$.  Once $T < T_c$,
the field $\varphi$ will roll or tunnel to its vacuum value.  In
doing so, $\varphi$ will oscillate about its {\sc vev} and radiate its
energy in neutrinos.  These neutrinos will carry a typical
energy of order $m_\varphi \simeq \vev{\varphi}$.

To estimate the potential signal, we first note that the total
energy stored in the symmetric phase $E_\varphi$ is given by
$\sim\langle \varphi\rangle^4\, V$; $V$ is the volume of the
neutrino-sphere. Comparing $E_\varphi$ to $E_\mathrm{SN}$ for
$\langle\varphi\rangle=1\,\mathrm{MeV}$, we see that only about
$10^{-6}\, E_\mathrm{SN}$ will be emitted directly during the phase
transition. To have a chance to detect such a tiny neutrino
emission, only the largest available detector with a very
low threshold can be used. The IceCube detector has excellent
capabilities to study supernova neutrinos, although its single
event energy threshold is around $100\,\mathrm{GeV}$. Instead,
supernova neutrinos will be detected only by the ice glow their
charged current interactions produce, which leads to an increase
in the noise rate of the photomultiplier tubes~\cite{Halzen}. The
signal in IceCube is thus proportional to $E_\nu^3$, where a
factor $E_\nu^2$ is due to the energy dependence of the cross
section and an additional factor $E_\nu$ accounts for the fact
that the number of \v{C}erenkov photons is directly proportional
to $E_\nu$. The number of neutrinos emitted in the transition
scales as $\vev{\varphi}^3$.  Since
$E_\nu\simeq\langle\varphi\rangle$, the total signal scales like
$\langle\varphi\rangle^6$. On the other hand, the minimum signal
strength needed for a significant detection depends on the noise
of the photo-tubes $f$ and on the time interval of the neutrino
emission $\Delta t_\varphi$ like $1/\sqrt{f\, \Delta t_\varphi}$.
Using a realistic set of numbers to describe IceCube from
Ref.~\cite{Halzen}, we find that a minimal signal
for the phase transition requires $\langle\varphi\rangle\geq 5
\,\mathrm{MeV}$ and $\Delta t_\varphi\leq 10^{-4}\,\mathrm{s}$.
This would require that the usual upper bounds on
$\langle\varphi\rangle$ be somewhat relaxed and also that the
whole neutrino-sphere undergo the phase transition at the same
time, in a {\it neutrino-flash}.

Next, we will present our quantitative results for the expected
size of the signal from $\varphi$-gas decays at the
neutrino-sphere. In thermal equilibrium, $\varphi$'s will track
the neutrino density and temperature closely, {\it ie.} they are
confined within the neutrino-sphere by the
$\varphi\leftrightarrow\nu\nu$ reactions. Neutrinos escape from
the surface of the neutrino-sphere. Similarly, $\varphi$'s also
escape, but decay into neutrinos promptly, as they have a lifetime
$\tau \sim T/(\lambda m_\nu)^2 \sim 10^{-6}\,\mathrm{s}$, in the
star's rest-frame. Thus, the neutrino-sphere will radiate
Fermi-Dirac thermal neutrinos, as well as those from decays of the
Bose-Einstein $\varphi$-gas.

The relative luminosity is then determined by the energy fractions
carried by neutrinos and $\varphi$'s. The two-component gas of
neutrinos and $\varphi$'s will have its energy equipartitioned
into all available degrees of freedom: 3 fermionic  and 1 bosonic.
Thus, the total number of effectively massless degrees of freedom
is $3\times(7/8)+1=29/8$. Hence, a fraction $F_B=8/29 \simeq 0.28$
of the total neutrino luminosity will be emitted from a
Bose-distributed gas. As the most important detection channel is
inverse $\beta$-decay, the next question is: how large is the
$\varphi$-decay contribution to the ${\bar \nu}_e$ flux?

The decay of the $\varphi$ particles produces mass eigenstates
instead of flavor eigenstates, hence the branching ratio into a
flavor $\alpha$ depends on the hierarchy of masses $m_i$
and the relevant mixing matrix elements $U_{\alpha i}\,$:
\begin{equation}
\label{eq:branching}
\mathrm{Br}({\nu_\alpha})\propto \sum_{i} |U_{\alpha i}|^2 m_i^2\,.
\end{equation}
Here, we note that neutrino oscillation data have provided some
information on $U_{\alpha i}$~\cite{Maltoni:2004ei}:
$|U_{e1}|^2\simeq0.7$ and
$|U_{e2}|^2\simeq0.3$ whereas for $|U_{e3}|^2$ there is only an
upper-bound of  $0.05$ at  the $3\,\sigma$ level. Since
$\mathrm{Br}({\nu_\alpha})$ is proportional to $m_i^2$,
we can distinguish three cases. The first one is
normal hierarchy, {\it ie.} $m_3$ is much heavier than $m_1$ and
$m_2$. In this case basically only the state $m_3$ is produced and
$\mathrm{Br}({\bar{\nu}_e})=1\times|U_{e3}|^2\leq0.05$. In the
case of inverted hierarchy, $m_1$ and $m_2$ have nearly the same
mass and are much heavier than $m_3$, which gives
$\mathrm{Br}({\bar{\nu}_e})=(1/2)\times|U_{e1}|^2+(1/2)\times|U_{e2}|^2\simeq1/2$.
In the case of degenerate neutrinos, all three masses are
approximately the same and
$\mathrm{Br}({\bar{\nu}_e})=(1/3)\times|U_{e1}|^2+(1/3)
\times|U_{e2}|^2+(1/3)\times|U_{e3}|^2\simeq1/3$. The branching
ratio of the $\varphi$-decay determines the flavor ratio for the
bosonic flux component:
$\eta^B_{\nu_\alpha}=\mathrm{Br}(\nu_\alpha)\,.$

In a simple blackbody approximation to the neutrino emission from
the proto-neutron star, there would be again equipartition and
thus the flavor ratio for the Fermi-Dirac component would be
$\eta^D_{\nu_\alpha}=1/3$.  Certainly, the effect in $\bar\nu_e$
would be very small for normal hierarchy  in the absence of
neutrino oscillations. More detailed studies of neutrino emission
including various levels of microphysical detail show a more or
less pronounced difference from the blackbody Ansatz and
deviations in $\eta^D_{\nu_\alpha}$ from $1/3$ up to a factor of
two seem possible~\cite{Keil:2002in}.

In general, the $\bar\nu_e$ flux arriving at the detector
$\Phi_{\bar\nu_e}$ is a
sum of 6 different initial fluxes at
the supernova $\phi_{\bar\nu_\alpha}$
\beq
\label{eq:fluxI}
 \Phi_{\bar\nu_e}=N\times\sum_{\bar\nu_\alpha}
 P_{\bar\alpha\bar e}\left(
F_D\eta^D_{\bar\nu_\alpha}\phi^D_{\bar\nu_\alpha}+
F_B\eta^B_{\bar\nu_\alpha}\phi^B_{\bar\nu_\alpha}
 \right)\,,
\eeq
where $P_{\bar\alpha\bar e}$ denotes the probability of neutrinos
of flavor $\bar\alpha$ to arrive as $\bar\nu_e$ at the detector.
$N$ is a normalization factor accounting for the energy and
distance of the supernova. Here, there is basically no
difference between $\bar\nu_\mu$ and $\bar\nu_\tau$. Both interact
with the neutron star material only via neutral current
interaction and any differences in their
initial spectrum would be eliminated by neutrino oscillation
via the atmospheric angle. Thus, it is sufficient to treat
$\bar\nu_\mu$ and $\bar\nu_\tau$ as one flavor $\bar\nu_x$
with twice the flux. Having now an effective two
flavor problem we can write $P_{\bar x \bar e}$ as $1-P_{\bar e
\bar e}$. Since $F_B+F_D=1$, Eq.~(\ref{eq:fluxI}) yields
\begin{eqnarray}
\label{eq:flux}
 \Phi_{\bar\nu_e}&=&N\times\left\{ P_{\bar e\bar e}\left[
(1-F_B)\eta^D_{\bar\nu_e}\phi^D_{\bar\nu_e}+
F_B\eta^B_{\bar\nu_e}\phi^B_{\bar\nu_e}
 \right]\right.\nonumber\\
 &+&\left.(1-P_{\bar e \bar e})\left[
(1-F_B)\eta^D_{\bar\nu_x}\phi^D_{\bar\nu_x}+
F_B\eta^B_{\bar\nu_x}\phi^B_{\bar\nu_x}
\right]\right\} \,.
\end{eqnarray}
The $\eta^B$'s are determined by the solar mixing angle and the mass
ordering. Therefore, we will discuss the different possibilities
for the mass ordering.  However, within each case, we will assume
each $\eta^B$ to be known, since the solar mixing angle is well
constrained.  $P_{\bar e \bar e}$
also depends on the solar mixing angle, the mass ordering,
and additionally on the small angle $\theta_{13}$.  The value of
$P_{\bar e \bar e}$ is determined by the adiabaticity of the
neutrino evolution at the MSW resonance which corresponds to the
larger, atmospheric mass splitting. The adiabaticity itself depends
on the value of $\theta_{13}$; for a detailed derivation see
{\it eg.}~\cite{Dighe:1999bi}.

Since there is only an upper-bound on $\theta_{13}$, we will
consider two extreme cases: either $\sin^2\theta_{13}$ is much
smaller than $10^{-3}$ or much larger. In the first case, $P_{\bar
e \bar e}=\cos^2\theta_{12} \simeq 0.3$, irrespective of the mass
ordering; in the second case $P_{\bar e \bar e}=\cos^2\theta_{12}$
for normal mass hierarchy and $P_{\bar e \bar e}=0$ for inverted
hierarchy. Thus, at least a fraction
$1-\cos^2\theta_{12}\simeq0.7$ of the $\bar\nu_x$ flux will
contribute to the observable ${\bar \nu}_e$ flux. Therefore,
there is always a sizable contribution to the signal
by neutrinos from $\varphi$-decay. Even for normal hierarchy, where
nearly all $\varphi$'s decay into $\bar\nu_x$, since at least $70\%$ of
$\bar\nu_x$ will oscillate into $\bar\nu_e$.

The spectral distribution for the initial fluxes in the
Fermi-Dirac case is given by the usual expression.  However, for
the neutrinos originating in $\varphi$-decays, we have to consider
that each $\varphi$ particle will decay into two neutrinos, with
flat energy distributions in the star's rest-frame. Hence, the
distribution functions we use are \beq
\phi^D_{\bar\nu_\alpha}(E; T_{\bar\alpha})=
\frac{E^2}{e^{E/T_{\bar\alpha}}+1}
\eeq
and
\beq \phi^B_{\bar\nu_\alpha}(E; T_{\bar\alpha})=
2\int_E^\infty \frac{d E^\prime}{E^\prime}\left(\frac{E^{\prime 2}}
{e^{E^\prime/T_{\bar\alpha}}-1}\right)\,,
\eeq
where the factor of $2$ in
$\phi^B$ accounts for 2 neutrinos per decay. We implicitly assume
that the temperature of the $\varphi$-gas and the Fermi-Dirac gas
are the same, but we allow for different temperatures of
$\bar\nu_e$ and $\bar\nu_x$. The signature for the presence of a
$\varphi$-gas is thus the observation of up to four different
distributions which compose the signal. Two of them are the usual
Fermi-Dirac distributed $\bar\nu_e$ and $\bar\nu_x$ contributions,
and we have another two contributions which result from
$\varphi$-decay. The latter two have an average energy which is
approximately only $1/2$ of that corresponding to the supernova
temperature.

Our simple Ansatz does not include possible changes in the
spectral shape with respect to the pure Fermi-Dirac or Bose
distributions. Those changes arise due to the energy dependent
position of the neutrino-sphere and due to diffusion processes on
the way out from the neutron star. Since neutrino density is
relatively low outside the neutrino-sphere,
Pauli-blocking can be ignored~\cite{Keil:2002in}.  Therefore, one
naively expects that the Bose and Fermi-Dirac components do not
interact on their way out. Thus, transport or diffusion should
affect both components in a similar fashion. To understand the
effects of the $\varphi$-gas on the position of the
neutrino-sphere(s) and its energy and flavor dependence requires a
detailed calculation along the lines of Ref.~\cite{Keil:2002in} and is
beyond the scope of this Letter. However, it seems unlikely that
such effects could conspire to destroy the four-component feature.

It remains to estimate the size of the event sample needed to
detect the bi-modal nature of the $\bar\nu_e$ spectrum. To this
end, we convert the flux in Eq.~(\ref{eq:flux}) to an
event rate by convolving it with the cross section for inverse
$\beta$-decay which is approximately given by
$\sigma=9.52\times10^{-44}\,E_\nu^2\,\mathrm{cm}^2\,\mathrm{MeV}^{-2}\,,$
with a $Q$-value of $1.8\,\mathrm{MeV}$. We can now compute the
event spectrum for a given set of parameters
$T_{\bar\nu_e},T_{\nu_x},\eta^D_{\bar\nu_e},P_{\bar e \bar e},F_B$
and ask how well we can determine or constrain $F_B$ from a fit to
these simulated data. For the actual computation we will use the
inverse $\beta$-decay cross section from Ref.~\cite{Vogel:1999zy}
and assume a typical detector threshold of $7\,\mathrm{MeV}$~\cite{Fukuda:2001nj}.
We use the supernova input parameters
\beq
T_{\bar\nu_e}=4.8\,\mathrm{MeV},\,\,T_{\nu_x}=4.8\,
\mathrm{MeV},\,\, \eta^D_{\bar\nu_e}=0.66\,,
\eeq
corresponding to the simulation in Ref.~\cite{Keil:2002in}. With $P_{\bar
e \bar e}=\cos^2\theta_{12}$, a supernova with a total energy
release of $3\times10^{53}\,\mathrm{erg}$ at a distance of
$10\,\mathrm{kpc}$, yields $\sim10\,000$ events, in a
detector with a fiducial mass of $22.5\,\mathrm{kt}$, like
Super-Kamiokande. We simulate our data assuming that indeed there
is a $\varphi$-gas, {\it ie.} $F_B=0.28$.

We perform a least square fit to the simulated data in the usual
way by constructing a $\chi^2$-function and asking how many events
are needed to exclude the absence of the $\varphi$-gas, $F_B=0$,
at $3\,\sigma$ confidence level.  More technically, we want to know
how large $\Delta\chi^2(F_B=0)$ is per event. The fit is constrained to
$T_{\bar\nu_e},T_{{\bar \nu}_x} \geq 3.8\,\mathrm{MeV}$,
which is a conservative
range~\cite{Keil:2002in}. The number of events needed for a
$3\,\sigma$ discovery, under various assumptions about our
knowledge of the supernova, is shown in the table.  From
this table, we see that Super-Kamiokande is quite capable of
detecting the presence of the $\varphi$-gas without the need for a
detailed model of the supernova physics.

Establishing the Bose-Einstein origin of the flux from
$\varphi$-decay will require a large data sample. Here, one can
perform an analysis similar to the one above, using a Fermi-Dirac
distribution for the scalar gas when testing the $\chi^2$
difference in the fit to the simulated data.  We find that $10^5 -
10^6$ events are needed to reach a conclusion at $3\,\sigma$.  For
this purpose, our estimate suggests, one of the megaton water
\v{C}erenkov detectors will be needed.
\vspace*{1ex}
\begin{center}
\begin{tabular}{l||c||c|c}
Hierarchy&Normal&\multicolumn{2}{c}{Inverted}\\
\hline
$P_{\bar e \bar e}$&$\cos^2\theta_{12}$&$\cos^2\theta_{12}$&0\\
\hline\hline
Perfect knowledge&538&152&294\\
$N$ unknown&1107&678&682\\
$N$ and $T_{\bar\nu_e}$ unknown&1157&709&682\\
$N,T_{\bar\nu_e}$ and $T_{\bar\nu_x}$ unknown&10502&7397&709\\
$N,T_{\bar\nu_e},T_{\bar\nu_x}$ and $\eta^D_{\bar\nu_e}$
unknown&13449&7432&709\\
\end{tabular}
\end{center}
\vspace*{1ex}

In this Letter, we presented supernova signatures of neutrino mass
models where new massive scalars with {\sc vev}'s 
in the range keV--MeV generate
$m_\nu \neq 0$.  We showed that the scalars will typically come to
thermal equilibrium during the initial stages of supernova
cooling. The subsequent decay of the scalars yields a neutrino flux
with roughly half the average energy of
thermally emitted neutrinos.  The lower energy component of the
neutrino flux encodes the Bose distribution of the parent
scalars.  A bi-modal energy distribution for each flavor is a
distinct and robust signature of the models we consider. This
effect could be readily detectable at the $3\,\sigma$ level at
Super-Kamiokande, for a typical galactic supernova.  Observation
of these signatures may offer the only direct
access to the physics responsible for neutrino mass generation.

%, within
%the class of models analyzed here.

\acknowledgments

It is a pleasure to thank V. Barger, A. de Gouv\^{e}a,
and H.-T. Janka for discussions.  We also thank G. Perez
and T. Schwetz for comments on the flux spectrum.
This work was supported in part by the United States Department of
Energy under Grant Contract No. DE-FG02-95ER40896.  H.D. was also
supported in part by the P.A.M. Dirac Fellowship, awarded by the
Department of Physics at the University of Wisconsin-Madison.

%%%%%%%%%%%%%%%%%%%%%%%%%%%%%%%%%%%%%%%%%%%%%%%%%%%%%%%%%%%%%%%%%%%%%%%%%%%

\end{document}